# HE-LHC: REQUIREMENTS FROM BEAM VACUUM

J.M. Jimenez, CERN, Geneva, Switzerland

*Abstract*

First thoughts on the design of the beam vacuum system for the High Energy LHC (HE-LHC) are given with a particular focus on the impact of the synchrotron radiation. In the HE-LHC, the vacuum dynamic effects induced by the circulating beams are expected to be as compared to the LHC. These effects will be reviewed and first thoughts on how to avoid or mitigate their effects are discussed.

## MACHINE PARAMETERS IMPACTING BEAM VACUUM

Even though the overall vacuum layout and integration issues could be very similar to the LHC [1], the parameter list of the HE-LHC [2] shows several changes, compared to the LHC, which can significantly affect the beam vacuum performances and stability [3].

The increase of the beam energy, of the bunch population, of the synchrotron radiation power and of the critical photon energy will influence the beam-induced effects taking place in vacuum systems which are linked both to the total intensity and to the bunched structure of the beams.

The decrease of the total number of circulating bunches, from 2808 to 1404 [2], will reduce the beam-induced effects in vacuum linked to the total beam intensity and will partly compensate the increase of the bunch population for the effects linked to the bunched structure of beams.

Finally, the increase of the beam potential resulting from the increase of the bunch population and emittance reduction, combined with the reduction of the magnet aperture will impact on the vacuum stability and electron cloud build up.

### Desorption induced by primary beam losses

The sections at cryogenic temperature are the most critical due to the potentially large quantities of condensed gasses which can be released resulting from a local heat load. However, these sections are "protected" by the quench limit of the cryomagnets. Indeed, the cryomagnets quench level [4] i.e. the number of lost protons to create a transition to the normal state, correspond to a negligible pressure rise ($<<10^{-6}$ Pa). Primary beam losses will induce a local desorption of gasses but would never lead to a vacuum limitation.

### Primary ionisation with circulating beams

The primary ionisation of the residual gas induced by the beams is linearly dependent on the ionisation cross section (about constant) and on the total intensity. As the ionisation cross-section is not expected to vary significantly between 7 and 16.5 TeV and taking into account the lower total intensity (60% of LHC), a similar effect as in the LHC is expected.

### Ion induced instability

The ion induced instability is linearly dependent on the desorption yield (about constant), on the ionisation cross-section (also approximately constant), and on the total intensity (0.6 times smaller) and is inversely proportional to the effective pumping speed. The later become the dominant factor for the vacuum stability. To ensure the vacuum stability along the sections at cryogenic temperature, only the pumping speed available through the beam screen pumping slots is considered. Then, considering the new beampipe aperture, the transparency of the beam screen shall be increased to 6.2% (as compared to the 4.4 % of the LHC), which could imply impedance and HOM issues. This issue has still to be addressed.

### Synchrotron radiation power

The synchrotron radiation power is proportional to the $4^{th}$ power of the energy and to the total beam intensity. An increase by a factor 17.3 is expected as compared to the LHC.

In the LHC, this heat load is intercepted by the beam screen. To keep such a design, an evaluation has to be made to ensure that the existing size of the cooling capillaries will be large enough to provide the cooling required. Any increase of the diameter of the capillary would lead to a further beam aperture reduction. An alternative could be to install photon absorbers in the cryomagnet interconnecting bellows (plug-in-modules), which would intercept the heat load outside the cryomagnets, in order to minimise the heat deposition onto the beam screens. The residual fraction of heat deposited on the beam screen would be determined by the length, aperture and bending angle of the dipole cryomagnets.

### Linear photon flux

The photon flux per unit length depends linearly on the beam energy and intensity. This flux is 30% higher than in the nominal LHC. Similarly to the LHC, a sawtooth structure shall be used in the beam screen to reduce the photon reflection and the photo-electron yield.

### Photon stimulated pressure rise

As compared to the LHC, the photon stimulated pressure rise is increased by a factor 7.4 since it grows with the $3^{rd}$ power of the beam energy and linearly with the beam intensity. This large increase is of concern for the vacuum system. Indeed, to ensure pressure stability, the pumping should be increased by the same amount which would bring the equivalent transparency of the beam screen to 46%! As this transparency would probably not be compatible with impedance and HOM issues, the vacuum system will have to rely of the vacuum cleaning i.e. reduction of the desorption yield ($\eta$). Details studies shall be launched to estimate the duration of the vacuum cleaning and confirm that it stay compatible with the operation constraints.

*Effects linked to the bunched structure of beams*

The electron and ion cloud build-up are two avalanche phenomena which can take place in the beam pipe. Both are threshold effects i.e. only take place above a given bunch population. As compared to the LHC, the bunch population has been increased by 12%, $1.29 \times 10^{11}$ p/bunch, well above the electron cloud threshold measured in the SPS i.e. $3.0 \times 10^{10}$ p/bunch in a dipole field [5]. The beam potential has also been increased by 30% resulting from the smaller longitudinal and transverse emittances. Based on these new parameters, an electron cloud build up can be expected. However, the reduction of the number of bunches by a factor 2 and the resulting bunch spacing of 50 ns has shown its efficiency to reduce the electron cloud build up, e.g. a reduction by a factor 10, as measured in the SPS.

Two other parameters playing a major role in the electron build up are varying: the beam screen height is decreased from 36.8 to 26 mm and the magnetic field is increased by a factor 2.4. Changing the beam screen aperture could bring the system out of resonance conditions. Indeed, increasing the beam potential will increase the energy of the primaries and finally, the small Larmor radius (few micrometers for a 100 eV electron) can also change the SEY yield. Simulations have to be done to provide information on the electron cloud build-up i.e. threshold and saturation levels.

As the beam will ionise the residual gas and due to the slow motion of the ions and enhanced by the secondary ionisation effect by the trapped electrons from the cloud (if any), an ion-induced positive space charge can take place. This phenomenon opens the risk for feedback effects. However, the reduction of the beam pipe aperture will probably cancel this effect.

*Feedback effects*

In presence of an electron cloud, part of the electrons can be trapped by an ion space charge. These electrons will spiral along the magnetic field and contribute to an additional ionisation of the residual gas. This secondary ionisation effect can lead to ion instability. This effect still needs to be quantified.

*Cold bore and beam screen operating temperature*

To ensure a proper pumping of hydrogen, the dominant residual gas in the beam vacuum, an operating temperature for the cryomagnets below 2-3 K is recommended. At higher temperatures, the hydrogen released will condensed up to an equivalent of a monolayer and then, the equilibrium pressure (hydrogen partial pressure) will start increasing very fast with the temperature i.e. $10^{-9}$ Pa at 2 K and up to $10^{-4}$ Pa at 4.2 K [6]. Similarly to what was made in the LHC, a beam screen will be required to shield the condensed gasses on the cold bore from the beam induced effects (electrons, ions and photon-stimulated desorption). Above 2-3 K, the use of cryosorbers will be required to ensure the required hydrogen pumping speed and capacity. The option of an operating temperature of the beams screen between 85 and 100 K can also be studied.

A major obstacle to increase the operating temperature of the beam screen from 5-20 K to 85-100 K could be the unacceptable increase of the magneto-resistance of the beam screen. This issue shall be investigated.

# REMEDIES TO VACUUM DYNAMIC EFFECTS

*Synchrotron radiation*

As made for the LHC, the use of a beam screen is required to intercept the synchrotron radiation induced heat load at a higher temperature. The use of photon absorbers will be considered, depending on magnet strength and length. At this stage of the discussion, the feasibility is not guaranteed. If considered, the cooling of these absorbers shall be decoupled from the cooling of the beam screens to preserve the cooling capacity of the beam screens. Similarly to what was done in the LHC, the photo-electron and photon reflection yields shall be reduced by using a sawtooth structure.

The photon and photo-electrons induced gas desorption will improve with time resulting from the vacuum cleaning effect (dose effect).

*Ion induced instability*

The design of the beam vacuum system shall be made to provide enough effective pumping speed considering beam pipe conductance. Considering the smaller aperture in the HE-LHC and the distributed induced gas desorption, the pumping provided by the pumping slots of the beam screen will dominate. The operating temperature of the cryomagnets is a key factor. As mentioned earlier, deeper calculations shall be made since the required transparency resulting from the preliminary estimations (46%) is certainly incompatible with impedance and HOM issues.

*Electron cloud suppression or mitigations*

The electron cloud is a fast avalanche and threshold phenomenon which behaviour depends on beam parameters. In existing machines, mitigation solutions are preferred since suppressing techniques cannot be easily retrofitted in an existing design.

For a new design, the suppressing techniques, e.g. techniques which prevent the electron avalanche to take place, shall be preferred. This will prevent any limitation for the future accelerator.

The suppressing techniques are often active solutions and the most commonly used are the clearing electrodes [7]. The use of clearing electrodes has many advantages since the electrodes capture the electrons right after their emission preventing any kind of avalanche effect. As compared to other solutions, this solution is not affected by venting to air and its efficiency is similar at ambient and cryogenic temperatures.

However their installation is complex since the clearing electrodes shall be in the vertical plane in the dipoles since electrons are confined along the dipole field lines. In the dipoles, the clearing electrodes shall be wide enough to cover the spacing of the vertical electron strips which spacing varies with bunch intensity.

An option for design could be to use the pumping port shields placed behind the pumping holes of the beam screens. Indeed, following the measurements made in the SPS, the LHC beam screens were equipped with shielding baffles placed between the beam screens and the cold bores and attached to the cooling capillaries. These baffles aim to intercept the electrons from the cloud, escaping from the beam screens through the pumping slots, to prevent the heat deposition onto the cold bore. Right from the design stage, the same configuration can be modified to convert the shielding baffles into clearing electrodes by insulating them from the cooling capillaries and polarising them to about 1 kV.

Coatings with a low secondary electron yield (SEY) are also mitigation solution to be considered. The coatings efficiencies depend on their ultimate SEY as compared with the needs of the accelerator.

Amorphous carbon coating is being considered in the SPS as LHC injector since it provides a low SEY (1.1) which is not affected by the venting to atmosphere. The behaviour of the amorphous carbon at cryogenic temperature will be investigated as an option for the sections operated at cryogenic temperatures. Another option is the NEG (TiZrV) coatings which also showed low SEY (1.1) after activation above 180°C. The need for a bake-out prevents its use in the sections at cryogenic temperature.

### Scrubbing Runs

The scrubbing runs aim to reduce the desorption yields ($\eta$) and the SEY ($\delta$) and to increase the bunch population threshold required to trigger an electron avalanche. This scrubbing effect is efficient only up to the bunch intensity used during the scrubbing periods, for a given filling pattern. Recent LHC studies with beams have confirmed the huge impact of the bunch spacing and length of bunch trains on the electron cloud build-up [8].

Measurements made in laboratories and observations on running accelerators have confirmed the efficiency of the scrubbing runs to decrease the electron cloud build-up. However, during these periods, the detectors cannot take any data during the scrubbing run since saturated by the background induced by the beam-gas scattering.

## GAS LOAD ISSUES IN CRYOGENIC SECTIONS

Similarly to the LHC, the HE-LHC shall take into account thick gas coverage of the beam screens (BS) and the cold bores (CB) by atoms/molecules desorbed directly (beam losses) and indirectly (photons, electrons and ions). Indeed, this could lead to pressure oscillation and vacuum instabilities.

In practice, the expected coverage should not become a limiting factor since mitigation solutions exist. In case of thick gas coverage in the BS, it can be recycled by heating up to 80 K. The gas will be "flashed" towards the cold bore through the BS pumping holes. The conditions can be met during short technical stop (2-3 days) similarly to what is planned for the LHC.

In case of thick gas coverage in the CB, it can be recycled by warming-up to 80 K. The gas will be pumped away using mobile turbomolecular pumps. These conditions will be met in the LHC, once per year during the Christmas technical stop.

## CLOSING REMARKS

### Start-up scenario

An accelerator vacuum system cannot be designed for nominal performances as on day-one. Often, its design rely on vacuum cleaning (reduction of desorption yields $\eta$ by photon, electron and ion bombardments) and on beam scrubbing (reduction of the secondary electron yields $\delta$).

With bunched beams, two options are possible. The first option is to start the operation with the nominal number of bunches and progressively increase the intensity per bunch. This allows to benefit from the vacuum cleaning effects and therefore the effects linked to the bunched structure of beams (electron cloud and ion instability) are less limiting since stimulated desorption coefficients ($\eta$) would have decreased with time/dose before reaching bunch intensity thresholds for electron cloud. It is important to underline that the beam pipes with two circulating beams will behave differently.

The second option is to start the operation with the nominal bunch intensity and progressively increase the number of circulating bunches. This allows for higher luminosities with lower machine optimisation but all effects linked to the bunched structure of beams (electron cloud and ion instability) will be at their maximum. Using this scenario implies limitation for the operation since vacuum cleaning and beam scrubbing time will be required to improve the situation.

The LHC requires both a vacuum cleaning and scrubbing period but some constraints could slow down these improvements: background to the experiments, induced heat load to cryogenics and cryomagnet quench limits (beam-gas scattering) prevent operation with large electron cloud which should have lead to a faster vacuum cleaning and beam scrubbing.

Considering what was observed in other accelerator and in particular in the LHC, the HE-LHC shall go for more conservative design: effects linked to the bunched structure of beams shall be suppressed at the design stage. This will help reducing the background to detectors and will help if the beam scrubbing of surfaces at cryogenic temperatures and cold/warm transitions is slower than initially considered. It will definitely save the operation in case the accumulation of the low energy electrons with high reflectivity (survivals) compensates the reduction of the secondary electron yield (SEY). Indeed, the beam scrubbing no longer help, photo-electrons production will dominate (design issue i.e. will not be significantly improving with time/dose).

*Vacuum system design: preliminary considerations*

The design of the HE-LHC beam vacuum shall be stable on day-one against ion-instability, reduce the number of photo-electrons and rely on vacuum cleaning (decrease of $\eta_{ph}/\eta_{e^-}$) for gas desorption stimulated by synchrotron radiation and photo-electrons.

This design would imply the use of a beam screens but as compared to LHC, the following issues must be looked at:
- More pumping speed is required i.e. more pumping slots;
- Mechanical constraints: deformation with quench, impedance and HOMs;
- Cooling capillaries are required to cool down the beam screens
- Operating temperature of the beam screen between 85-100 K is being favored provided that the magnetoresistance of the beam screen stays compatible with impedance requirements;
- Cryosorbers are required in the cold bore side if the cryomagnets are operated above 3 K ;
- Clearing electrodes in dipoles behind the beam screens and attached to the cooling capillaries to suppress electron cloud, alternatively:
  • Proceed to a coating of quadrupoles and cold/warm transitions of standalone cryomagnets;
  • Use solenoids (3-5 mT) to mitigate electron cloud build up in vacuum instrumentation ports and interconnecting pieces which cannot be coated;
  • Long straight sections at ambient temperature should be baked and rely on NEG coatings, alternatively, install solenoids if the coating is not feasible.

These first thoughts on the design of the HE-LHC beam vacuum system need to be revisited once all pending issues have been correctly evaluated.

## ACKNOWLEDGEMENTS

The author would like to thank O. Grobner and V. Baglin for their helpful discussions.